\documentclass[fleqn,usenatbib]{mnras}

\usepackage{newtxtext,newtxmath}
\usepackage[T1]{fontenc}
\usepackage{graphicx}	
\usepackage{amsmath}
\usepackage{orcidlink}
\usepackage{multirow}
\usepackage{ltxtable}
\usepackage{threeparttable}
\usepackage{comment}
\usepackage{tikz}\usetikzlibrary{shapes.geometric, arrows}
\usepackage{pgfplots}\pgfplotsset{compat=1.18}
\usepackage{pifont}
\usepackage{tikz}
\usetikzlibrary{arrows.meta}
\tikzset{%
  >={Latex[width=2mm,length=2mm]},
    base/.style = {
        rectangle, rounded corners, draw=black,
        minimum width=2cm, minimum height=1cm,
        text centered, font=\sffamily
    }, 
    blind/.style = {base, fill=blue!30},
    intermediate/.style = {base, fill=orange!15},
    discarded/.style = {base, fill=red!60},
    verified/.style = {base, fill=green!60}
}

\DeclareRobustCommand{\VAN}[3]{#2}
\let\VANthebibliography\thebibliography
\def\thebibliography{\DeclareRobustCommand{\VAN}[3]{##3}\VANthebibliography}

\usepackage[utf8]{inputenc}
\usepackage{blindtext}
\usepackage{color}
\usepackage{everysel}
\usepackage{fancyhdr} 
\usepackage{float}
\usepackage{physics}
\usepackage{textcomp}
\usepackage{clrscode}
\usepackage{verbatim}
\usepackage[none]{hyphenat} 
\usepackage{amsfonts} 
\usepackage{caption}
\usepackage{subcaption}
\usepackage{siunitx}
\usepackage{mathtools}
\usepackage{lipsum}
\usepackage{tgcursor}
\usepackage{cases}
\usepackage{multirow}
\usepackage{ragged2e}

\usepackage{xcolor}
\usepackage{longtable}
\usepackage{booktabs}
\usepackage{graphicx} 
\usepackage{booktabs} 
\usepackage{xparse} 
\usepackage{supertabular}
\usepackage{makecell}

\usepackage{lscape}   
\usepackage{hyperref} 

\usepackage{array}
\newcolumntype{x}[1]{>{\centering\arraybackslash\hspace{0pt}}p{#1}}

\newif\ifsupp
\supptrue

\newif\ifbf
\bffalse

\usepackage{hyperref}
\hypersetup{
    colorlinks=true,
    linkcolor=blue,
    filecolor=magenta,      
    urlcolor=cyan,
}

\newcommand{\namedref}[2]{\hyperref[#2]{#1~\ref*{#2}}\xspace}

\newcommand{\sersic}{S\'{e}rsic\xspace}

\title[Bulges of Dual AGN Galaxies]{Investigating the Bulge Morphology of Dual AGN Host Galaxies from the GOTHIC survey }

\begin{document}

 \author[C.P.Nehal]{
 C.P.Nehal$^{1}$\thanks{E-mail : ncp.sudo@gmail.com}
 Mousumi Das\orcidlink{0000-0001-8996-6474},$^{2}$\thanks{email :  mousumi@iiap.res.in} 
 Sudhanshu Barway\orcidlink{0000-0002-3927-5402},$^{2}$
 Francoise Combes\orcidlink{0000-0003-2658-7893},$^{3}$ 
 Prerana Biswas\orcidlink{0009-0006-7375-6580},$^{2}$ 
 \newauthor
 Anwesh Bhattacharya,$^{4}$ 
 Snehanshu Saha$^{5}$\orcidlink{0000-0002-8458-604X},$^{4,5}$ 
 \\
 $^{1}$ Department of Physics, Indian Institute of Science Education and Research, Bhopal, India - 462066\\
 $^{2}$ Indian Institute of Astrophysics, Bangalore, India - 560034\\
 $^{3}$Observatoire de Paris, LERMA, College ` de France, PSL University, Sorbonne University, CNRS, Paris, France - 75014\\
 $^{4}$Siebel School of Computing and Data Science, University of Illinois at Urbana-Champaign, USA, IL-61801\\
 $^{5}$5APPCAIR, Department of CSIS, Birla Institute of Technology \& Science, Goa and HappyMonk AI, India - 403726
 }

\date{Accepted 2025 October 30. Received 2025 October 27; in original form 2025 September 19}

\pubyear{\the\year{}}


\label{firstpage}
\pagerange{\pageref{firstpage}--\pageref{lastpage}}
\maketitle

\begin{abstract}
We present a structural analysis of bulges in dual active galactic nuclei (AGN) host galaxies. Dual AGN arise in galaxy mergers where both supermassive black holes (SMBHs) are actively accreting. The AGN are typically embedded in compact bulges, which appear as luminous nuclei in optical images. Galaxy mergers can result in bulge growth, often via star formation. The bulges can be disky (pseudobulges), classical bulges, or belong to elliptical galaxies. Using SDSS DR18 \textit{gri} images and \textsc{GALFIT} modelling, we performed 2D decomposition for 131 dual AGN bulges (comprising 61 galaxy pairs and 3 galaxy triplets) identified in the GOTHIC survey. We derived \sersic indices, luminosities, masses, and scalelengths of the bulges. Most bulges (105/131) are classical, with \sersic indices lying between $n=2$ and $n=8$. Among these, 64\% are elliptical galaxies, while the remainder are classical bulges in disc galaxies. Only $\sim$20\% of the sample exhibit pseudobulges. Bulge masses span $1.5\times10^9$ to $1.4\times10^{12}\,M_\odot$, with the most massive systems being ellipticals. Galaxy type matching shows that elliptical--elliptical (E--E) and elliptical--disc (E--D) mergers dominate over disc--disc (D--D) mergers. At least one galaxy in two-thirds of the dual AGN systems is elliptical and only $\sim$30\% involve two disc galaxies. Although our sample is limited, our results suggest that dual AGN preferentially occur in evolved, red, quenched systems, that typically form via major mergers. They are predominantly hosted in classical bulges or elliptical galaxies rather than star-forming disc galaxies.
\end{abstract}

\begin{keywords}
 galaxies: active – galaxies: evolution – galaxies: interactions – galaxies: nuclei - galaxies: bulges – galaxies: structure.
\end{keywords}

\section{Introduction}
Galaxy mergers are important for the hierarchical growth of galaxies and the formation of structure in the universe \citep{white1991galaxy}. As the merging galaxies come closer, the gravitational torques generated by their interaction affects their stellar and gas distributions, resulting in increased star formation and the growth of stellar mass \citep{hopkins.etal.2010, Hopkins2009}. Studies show that gas-rich major mergers or wet mergers, can have star formation rates (SFRs) that are $\sim$100 times that before the interaction. Such enhanced SFRs are often found in starburst galaxies or ultraluminous infrared galaxies (ULIRGs) \citep{Mihos1996,Cox2004,li.etal.2025}. The merger-induced SFR is the highest for similar galaxy mass ratios, as observed in ULIRGs \citep{nandi.etal.2021}, and drops rapidly with increasing galaxy mass ratios \citep{Cox2008}. If the galaxies are gas-poor, then there is little or no associated star formation, and the merger is called a dry merger. The ensuing bulge or central spheroidal growth in the merger remnant arises from merger accreted stars from the companion galaxies \citep{quilley.lapparent.2022}, especially in the case of minor mergers \citep{zavala.etal.2012}.

The enhanced nuclear gas inflow and star formation activity can trigger mass accretion onto the supermassive black holes (SMBHs), in which case the nuclei become active galactic nuclei (AGN) \citep{schechter.etal.2025}. When both nuclear SMBHs start accreting mass they will form an AGN pair, or in other words a dual AGN \citep{rubinur.etal.2019,rubinur.etal.2021}. Although the initial detection of dual AGN was serendipitous \citep{komossa.2003}, recent surveys have found large samples of dual AGN \citep{Gothic2023,zhang.etal.2021}. A few triple AGN have also been detected \citep{keshri.etal.2025,yadav.etal.2021, pfeifle.etal.2019}. However, dual AGN are still rare and triple AGN even rarer. In the literature, dual AGN are generally defined as close mergers with AGN separations $\sim$1 to 40 kpc, whereas binary AGN have nuclei separations $<$100pc \citep[eg.][]{rubinur.etal.2018,deRosa.etal.2019}. Since galaxy mergers are often associated with star formation, the nature of nuclei in close mergers can have different combinations, such as star forming pairs (SF-SF), mixed pairs (SF-AGN) or dual AGN \citep{Gothic2023}. Since the discovery of gravitational waves from merging black holes \citep{abbott.etal.2016}, there has been a growing interest in detecting and studying dual and binary AGN \citep{mondal.etal.2024,bailes.etal.2021,kharb.etal.2017}. This is because inspiralling SMBHs at separations $<1$pc will give rise to nHz and mHz gravitational waves \citep{burke-spolaor.etal.2019}, and so AGN pairs are a means to study the formation of binary SMBHs \citep{padmanabhan.loeb.2024}. A good example of a binary SMBH system in a very late stage of merging is OJ287, but very little is known about its host galaxy \citep{valtonen.etal.2022,valtonen.etal.2025}.

During mergers, as the individual SMBHs spiral towards the center of mass, they will still be surrounded by a significant number of closely bound stars and thus be embedded in compact bulges or spheroids. In optical or X-ray images, these bulges will appear as bright nuclei \citep{das.etal.2018} and some may host dual AGN, depending on the SMBH accretion rates \citep{koss.etal.2012,giri.etal.2022}. Although dual AGN properties have been extensively studied using multiwavelength observations, not much is known about their host bulges. Are they all classical bulges that are spherical in shape or do they have oval or boxy morphology? For example, if the host galaxies are disk galaxies, the SMBHs maybe be embedded in bulges that are more disky, since their stellar mass may have grown via gas accreted along the galaxy plane during mergers \citep{zavala.etal.2012}. Alternatively, the bulges maybe boxy or peanut shaped if they evolved from bars instabilities or disk  thickening \citep {ghosh.etal.2024}. It is also possible that they have grown through a combination of both processes \citep{mendez-abreu.etal.2014}; both processes are part of the secular evolution of galaxy disks \citep{kormendy.kennicutt.2004}. 

In dry mergers, the merging galaxies are usually elliptical or lenticular, and the SMBHs will be embedded in extended spheroidal bulges. So understanding the bulge-disk morphology of large samples of merging galaxies can reveal the nature of the merging galaxies as well as predict the possible outcome of the merger, i.e. is an elliptical merging pair or a disk galaxy pair more likely to form a dual AGN? Or are dual AGN more likely to be found in mixed disk-elliptical merging pairs? Also important is that AGN are known to evolve with their bulges, and their co-evolution leads to the well known correlation between the nuclear velocity dispersion (or bulge luminosities), with SMBH mass ($M-\sigma$ relation) \citep{marsden.etal.2020}. Hence, deriving bulge morphologies in dual AGN is another way of understanding AGN-bulge co-evolution in different environments. Finally, bulge evolution in merging galaxies is an important part of the larger picture of galaxy evolution. Thus, there are a plethora of reasons for understanding the nature of bulges in dual AGN as well as in merging galaxies. 

Bulges are broadly of two types, classical bulges and pseudobulges \citep{fisher.drory.2016}. Classical bulges are dispersion-dominated stellar systems and appear as compact, bright spheroids in galaxy centers, whereas pseudobulges have a relatively stronger disk component and appear as oval or boxy in shape  \citep{debattista.etal.2004,kumar.etal.2021}. Classical bulges are formed in early epochs due to the monolithic collapse of gas clouds, and continue to grow through mergers \citep{eggen.etal.1962,brooks.etal.2016}, whereas pseudobulges are formed via secular evolution of galaxy disks \citep{kormendy.fisher2008}. Bulges can be quantitatively analyzed using the \sersic index value $n$ \citep{graham.driver.2005}. Pseudobulges have a \sersic index value $n<2$, while classical bulges have values $n>2$ \citep{gadotti.2009}. Studies show that bulges are closely related to their host galaxies; the disks of pseudobulge host galaxies are younger and have more star formation compared to the disks associated with classical bulges \citep{hu.eta.2024,vaghmare.etal.2013}. Also, the fraction of pseudobulges increases at lower redshifts as the bulge to disk fraction evolves with times \citep{kumar.kataria.2022}. In general the effect of mergers on bulges also depends on redshifts, as merger rates were higher at early epochs \citep{Bridge2010, sachdeva.etal.2017}. 

In this paper we focus on deriving the bulges and host galaxy types of dual AGN. Our main aims are the following : (i)~determine the \sersic indices of the DAGN bulges using the bulge-disk decomposition program GALFIT. (ii)~Hence, determine the nature of the host galaxies of the dual AGN. (iii)~Derive bulge masses and bulge mass ratios. (iv)~Determine whether dual AGN are more likey to be found in elliptical galaxy pairs, disk galaxy pairs or mixed elliptical-disk pairs. This is an important question as it addresses the probability of finding SMBH pairs in early type galaxies or disk galaxies. The paper is organized as follows. Section~2 describes the sample and outlines the two-dimensional decomposition of the galaxies. Section~3 presents the results of the decomposition. Section~4 discusses these results. A summary is provided in Section~5. Throughout this paper, a flat $\Lambda$CDM cosmology is used with $H_0 = 70\,\mathrm{km\,s^{-1}\,Mpc^{-1}}$, $\Omega_m = 0.3$, and $\Omega_\Lambda = 1 - \Omega_m$.

\section{Sample and Analysis}
\subsection{GOTHIC sample} 
Our sample is drawn from our previous study of galaxy mergers where we detected pairs of galaxy nuclei in an automated way using a novel algorithm called GOTHIC \citep{Gothic2023}. After applying the algorithm to a sample of one million galaxies derived from SDSS DR16 we obtained a confirmed sample of 681 close nuclei pairs. Furthermore, to classify the type of nuclear activity in the host galaxy nuclei, we applied the AGN diagnostic plot of Baldwin, Phillips, and Telverich (BPT) \citep{bpt.1981}. Using the BPT plot we derived a sample of 159 dual AGN (DAGN), which also included two triplet and one quadruplet system of galaxies. We used this sample of dual AGN in this study. he systems in this study are examples of binary galaxies which have evolved further into very close merging systems. How nuclear activity arises in the early stages of separate pairs of galaxies should have a bearing on the activity when the merger has happened, as discussed in \citet{byrd.valtonen.2001}.

\subsection{2-D image decomposition of galaxies}
Bulge–disc decomposition was carried out using the GALFIT software package \citep{peng.etal.2002} to investigate bulge properties in a sample of dual AGN. Imaging data were retrieved from SDSS DR18, which provides calibrated FITS images for the target galaxies. A 200x200 pixel cutout centred on one nucleus was extracted for each system. As the FITS images are in units of nanomaggies per pixel, a conversion to counts per pixel, required by GALFIT, was applied using the NMGY scaling factor provided in the FITS headers. Although SDSS DR18 provides sky-subtracted images with near-zero background levels, GALFIT requires a non-zero sky value for numerical stability. To satisfy this condition, a constant offset of 1000 counts was added to all pixels, and the same value was specified as the sky level in the GALFIT input. This modification significantly improved the convergence and robustness of the fitting procedure. A point spread function (PSF) image was constructed from a nearby unsaturated star in the same field and provided to GALFIT to account for PSF convolution during the modelling process. Foreground stars and background sources occasionally introduced poor fits and inflated reduced chi-squared ($\chi^2_\nu$) values. Mask images were created to exclude these objects and were incorporated into the GALFIT input via the feedme files. Galaxies with angular sizes comparable to the PSF were excluded to ensure reliable structural measurements. A minimum size criterion of twice the PSF full width at half maximum (FWHM) was adopted, resulting in a refined sample of 104 dual AGN systems from an initial set of 159.

\begin{figure}  
    \centering
    \begin{subfigure}[b]{0.30\columnwidth}
        \centering
        \includegraphics[width=\textwidth]{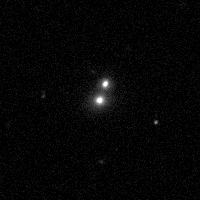}
        \caption{Image}
    \end{subfigure}
    \hfill
    \begin{subfigure}[b]{0.30\columnwidth}
        \centering
        \includegraphics[width=\textwidth]{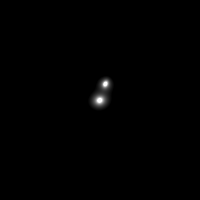}
        \caption{Model}
    \end{subfigure}
    \hfill
    \begin{subfigure}[b]{0.30\columnwidth}
        \centering
        \includegraphics[width=\textwidth]{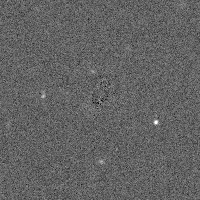}
        \caption{Residual}
    \end{subfigure}
    \caption{An example to show the result obtained from GALFIT (objid: 1237668271362211975)}
\end{figure}

Initial parameter estimates for all the sample galaxies were obtained from the SDSS 'photoObj' catalogue. To ensure reliable structural fitting, the PSF full width at half maximum (PSF-FWHM) was compared with the de Vaucouleurs radius (deVRad) in each of the five bands. Galaxies with deVRad < 2 x PSF-FWHM were excluded from the analysis. After applying this criterion, a final sample of 104 merging galaxies was selected for GALFIT modelling.

Estimation of initial parameters such as total magnitude, scale radius, axis ratio, and position angle was obtained from the 'photoObj' catalogue using the 'deVMag', 'deVRad', 'deVAB', and 'deVPhi' catalogue parameters, respectively, for each band. These values were used as input in the feedme files to fit a \sersic profile to the bulge component. In most cases, parameters were left free to vary during fitting to allow convergence on the optimal model. However, in a few instances, parameters were fixed to achieve a stable solution. Given that the initial guesses were close to the expected values, it was assumed that GALFIT would converge with minimal iterations. The \sersic index was initially set to 1 for all galaxies and allowed to vary during the fitting process. In cases where optical imaging suggested the presence of a disk, an exponential component was included in the model. An initial parameter set for the exponential disk was adopted and used in the fitting process when applicable. The corresponding parameters from the photoObj catalogue 'expMag', expRad, 'expAB', and 'expPhi' were provided as inputs in the 'feedme' file. GALFIT was initially executed for all sources to obtain an rms sky estimate, which was subsequently fixed as the sky background value in the corresponding feedme file. Fixing the sky parameter reduced the number of free parameters and, consequently, the number of fitting iterations. In cases with nearby contaminating sources, a custom mask was manually generated and supplied as 'mask.fits'. For some cases, for fitting purposes, an additional \sersic component was included to model their contribution when there is an extended source nearby.

As the sample consists of merging galaxies, each system contains multiple nuclei, necessitating a multi-component fitting approach. An initial fit was performed using one \sersic component per galaxy, each representing a bulge. Thus, every system includes at least two \sersic components. In cases where the fit was inadequate or a disc was visibly present, an additional exponential component was included to account for the disc. These sources, therefore, include both \sersic and disc components in the final model. 

GALFIT produced satisfactory fits, with acceptable $\chi^2/\nu$ values, for 69 out of 104 sources (Figure~1). For the remaining objects, poor residual images indicated model inadequacies. In several cases, GALFIT failed to converge and terminated upon reaching the maximum number of iterations, resulting in no output model. The fitting procedure employed a \sersic profile for the bulge and an exponential profile for the disc. However, certain disc morphologies exhibited complex features that this combination could not adequately capture. Several challenges were encountered during the fitting process, leading to the exclusion of a subset of sources from further analysis:

\begin{itemize}
\item In some cases, GALFIT did not converge owing to insufficient signal-to-noise in one or both galaxies, making extracting reliable structural parameters unfeasible. 
\item In other cases, the fitting process was not merely affected by contamination from neighbouring sources but failed significantly, resulting in spurious magnitudes and associated parameters. Moreover, systems exhibiting tidal disruption or pronounced morphological disturbances consistently produced unreliable fits, as such features deviate substantially from the assumptions inherent in axisymmetric models.
\item  For some galaxies, PSF information and initial fitting parameters from SDSS were unavailable, preventing the generation of suitable \texttt{galfit} feedme files. Where possible, manual estimates were used in conjunction with available PSFs, but these fits were found to be unreliable and the sources were subsequently excluded.  \item Several systems initially appeared as single sources but were classified as multiple components by SDSS, complicating the modelling.
\item In some cases, closely spaced nuclei led to confusion during fitting, with model functions attempting to fit neighbouring components, resulting in poor residuals.
\item  Two disk galaxies exhibited intersecting structures and highly diffuse, non-uniform light profiles, which could not be reliably fitted.
\end{itemize}

After excluding these problematic cases, the sample was reduced from 104 to 69 nuclear pairs used for further analysis. The output of GALFIT is total magnitudes, effective radii, and \sersic index for the bulge, along with total magnitudes and scale lengths for the disc in each band. The magnitudes in the g and r bands were used to obtain colours for the bulge and disc.

Magnitudes were corrected for Galactic extinction, and K-corrections were also applied to derive the (g–r) colours of bulges in the sample galaxies \citep{schlegel1998applicationsfddustmaps, 2010MNRAS.405.1409C, 2012MNRAS.419.1727C}. Five bulges exhibited unphysical negative (g–r) colours, inconsistent with the expected absence of young stellar populations in these systems. These were attributed to fitting errors in the magnitude estimates and were excluded from the analysis. The final sample comprises 64 reliable bulge pairs; this included 61 dual AGN systems and 3 confirmed triplets (including dual AGN), yielding a total of 131 bulges. The derived colours and applied corrections are listed in Table~\ref{table: color correction}. 

For one of the sources in the 61 pairs, there is a third source lying in between (objid 1237661812274233474). Although SDSS has classified this source as a GALAXY, there is no spectroscopic data available for the source, and hence no redshift. So we are unable to do further analysis for this source. However, this source had to be fitted in GALFIT in order to fit the neighbouring sources. Hence Table~\ref{table: color correction} contains this source with a footnote on why data columns are missing for the source. So the table contains 132 data rows while our study sample has 131 sources. 

\begin{figure}
	\includegraphics[width=8cm]{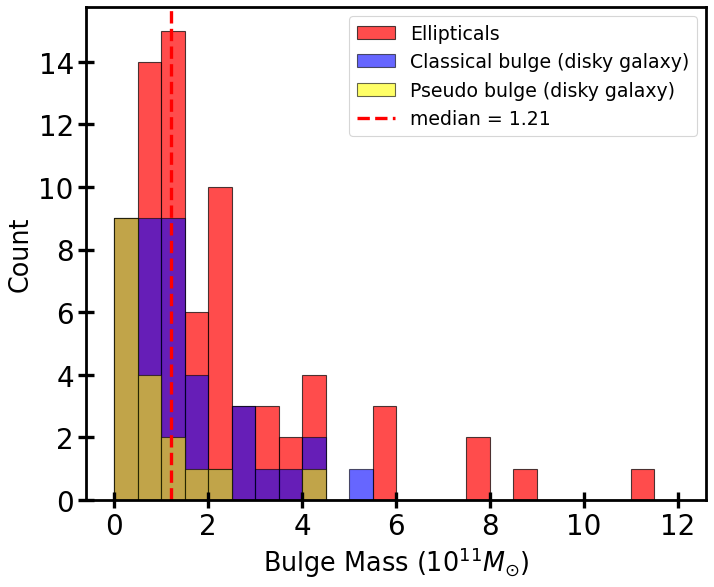}
    \caption{The bulge mass distribution of all the nuclei in the sample. Note that in the sample 73 were elliptical galaxies and had no disk, whereas 59 were disk galaxies with bulges.}
    \label{fig:bulge mass dist}
\end{figure}

\section{Results}
\begin{figure}
    \centering
\includegraphics[width=7cm]{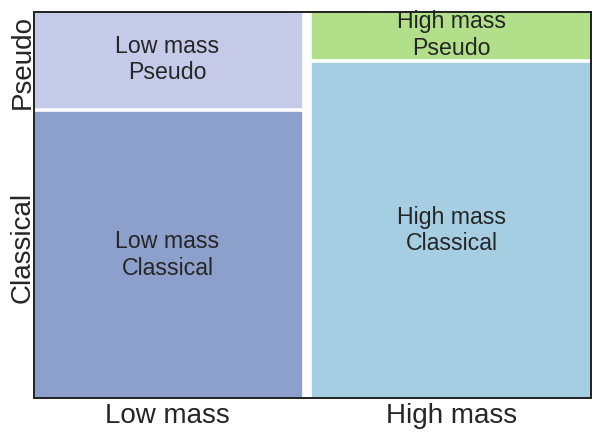}
    \caption{Mosaic of bulge mass and bulge type where bulge mass $1.21\times10^{11} \textup{M}_\odot$ separates the high and low mass bulges.}
    \label{fig:mosaic-bulge-disk}
\end{figure}

The $(g - r)$ colour and $i$-band magnitudes were used to estimate the mass-to-light (M/L) ratio and bulge mass following the prescription of \citet{bell2003}. The relation
\begin{equation}
\log_{10}(M/L) = a_{\lambda} + b_{\lambda} \times \text{(colour)}
\end{equation}
was applied using appropriate coefficients based on the chosen colour and photometric band. The resulting bulge masses are listed in Table~\ref{table: color correction}, and their distribution is shown in Fig.~\ref{fig:bulge mass dist}. Bulge masses span the range $0.015 \times 10^{11}$ to $1.376 \times 10^{12}~\text{M}_{\odot}$, with the majority falling below $2.5 \times 10^{11}~\text{M}_{\odot}$. One outlier, with $M = 1.10 \times 10^{12}~\text{M}_{\odot}$ (objID = 1237655692474515647), may correspond to an elliptical galaxy rather than a bulge in a disc galaxy; this possibility is discussed further below. A median bulge mass of $1.21 \times 10^{11}~\text{M}_{\odot}$ was adopted to separate the sample into low- and high-mass bulges. A mosaic of this division is presented in Fig.~\ref{fig:mosaic-bulge-disk}.

\begin{figure*}
    \centering
    \hspace{0.3cm}
    \begin{subfigure}[b]{0.30\textwidth}
        \centering
        \includegraphics[width=\textwidth]{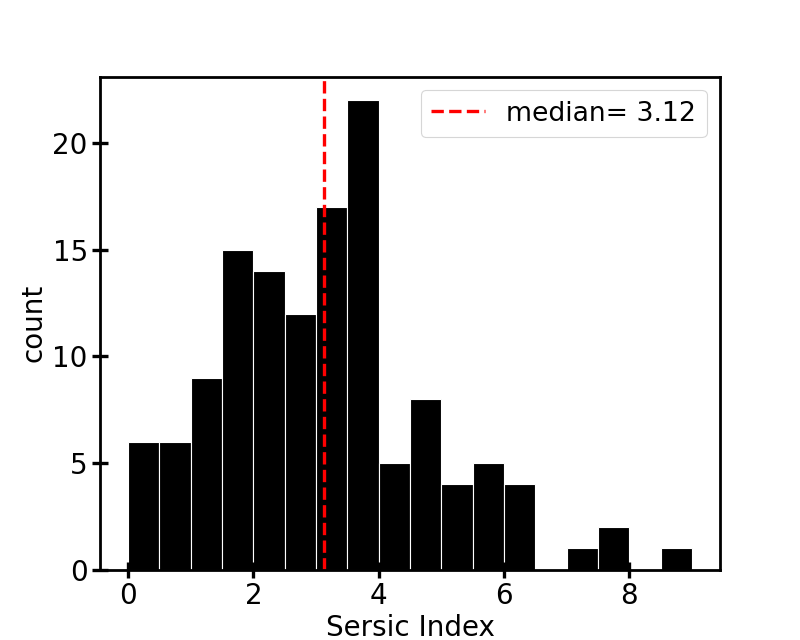}
        \caption{Sérsic Index distribution in g-band}
        \label{fig:g-band sersic hist}
    \end{subfigure}%
    \hspace{0.17cm}
    \begin{subfigure}[b]{0.30\textwidth}
        \centering
        \includegraphics[width=\textwidth]{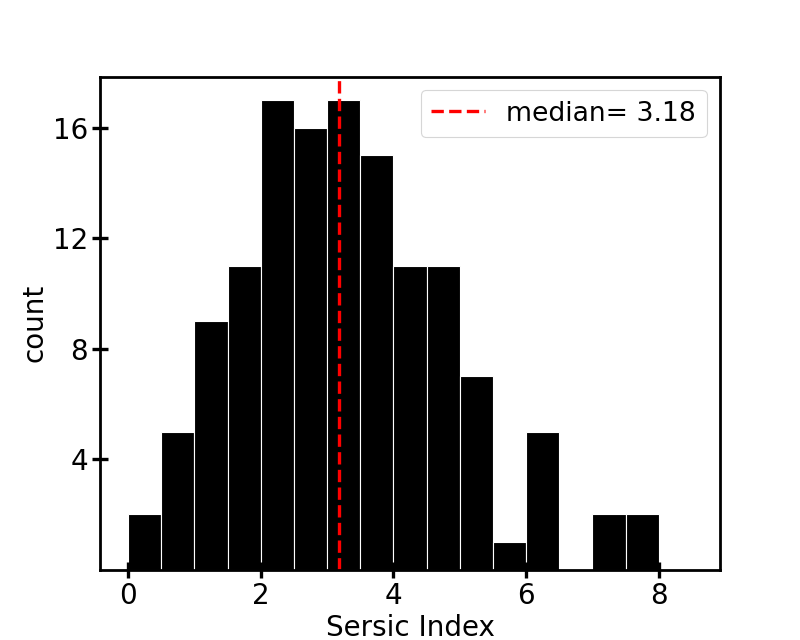}
        \caption{Sérsic Index distribution in r-band}
        \label{fig:r-band sersic hist}
    \end{subfigure}%
    \hspace{0.17cm}
    \begin{subfigure}[b]{0.30\textwidth}
        \centering
        \includegraphics[width=\textwidth]{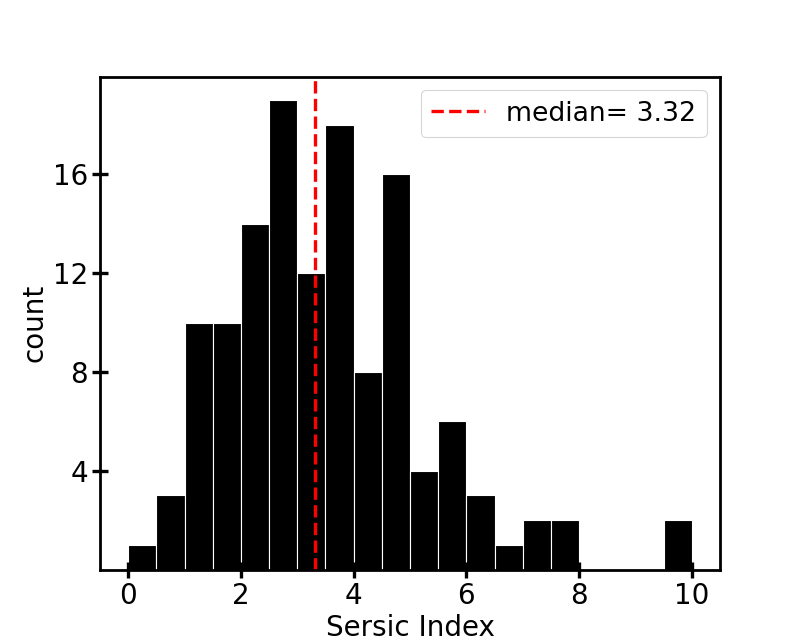}
        \caption{Sérsic Index distribution in i-band}
        \label{fig:i-band sersic hist}
    \end{subfigure}
        \caption{Distribution of Sérsic Index observed over three bands: 'g', 'r', and 'i' from left to right.}
    \label{fig:sersic_index_distribution}
\end{figure*}

The distribution of bulge \sersic indices in the g, r, and i bands is shown in Figure~\ref{fig:sersic_index_distribution}, using uniform bins of width 0.5. The index ranges and median values are as follows:
 \noindent
g-band: $0.15 \leq n \leq 8.56$, median = 3.13;
r-band: $0.33 \leq n \leq 7.87$, median = 3.17;
i-band: $0.43 \leq n \leq 9.95$, median = 3.27.
Adopting the conventional threshold of $n = 2$ to distinguish classical bulges ($n \geq 2$) from pseudobulges ($n < 2$; \citealt{gadotti.2009}), we find that 70–80\% of bulges in the sample have $n \geq 2$, indicating a predominance of classical bulges. The distributions are similar across all bands, with redder bands tending to yield slightly higher Sérsic indices. This consistency across bands suggests that the prevalence of classical bulges is not an artefact of band-dependent effects. We further examined the relation between bulge type and bulge mass (Figure~\ref{fig:mosaic-bulge-disk}), which shows that both the most massive bulges and a significant fraction of lower-mass bulges exhibit classical bulge nature.

The bulge mass and Sérsic index ($n$) serve as key morphological indicators. Two galaxy types are distinguished based on GALFIT decomposition: 

\begin{itemize}
    \item[(i)] \textbf{Disc galaxies with bulges}: As described in Section~2, if the GALFIT fitting required more than one Sérsic component, the second was consistently modeled as an exponential profile representing a disc. The bulge component is characterized by a Sérsic index $n$.
    
    \item[(ii)] \textbf{Elliptical galaxies}: These are well described by a single Sérsic component and lack an associated disc.
\end{itemize}

Table~\ref{table:properties} summarizes the combinations of structural components derived from GALFIT. Among the 131 bulges analyzed, 58 are associated with an exponential disc component and are classified as disc galaxies. The remaining 73 lack a disc and are classified as elliptical galaxies. The distribution of bulge and galaxy types is presented in Table~\ref{tab:bulge_disk}.

Of the 73 elliptical galaxies ($\sim 64\%$ of the sample), 67 host classical bulges and 6 exhibit pseudo-bulge profiles. These pseudo-bulge ellipticals may retain disky features due to past mergers, or they may be remnants of galaxies that lost their discs through tidal stripping during mergers. Among the 58 disc galaxies, classical bulges dominate, although a substantial number also host pseudo bulges. Overall, approximately $77\%$ of pseudo bulges are associated with disc galaxies, while only a small fraction reside in systems without discs.

\begin{figure*}
    \hspace{-0.25cm}
    \centering
    \begin{subfigure}[b]{0.28\textwidth}
        \centering
        \includegraphics[width=\textwidth]{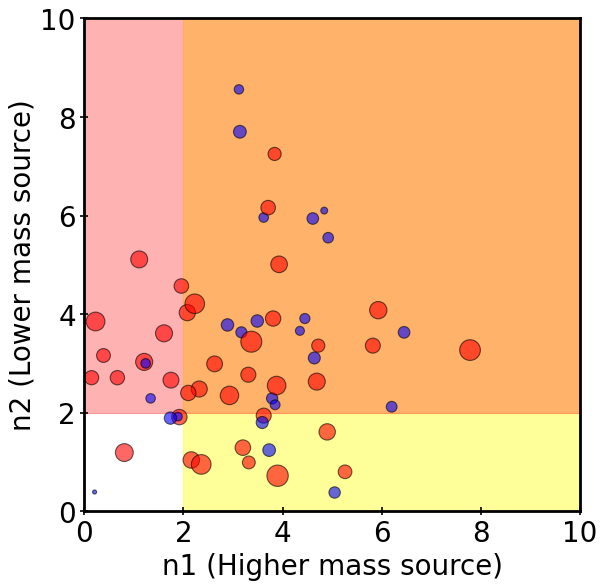}
        \caption{Combination of \sersic index in g-band}
        \label{fig:Sersic combination g-band}
    \end{subfigure}
    \hspace{0.4cm}
    \begin{subfigure}[b]{0.28\textwidth}
        \centering
        \includegraphics[width=\textwidth]{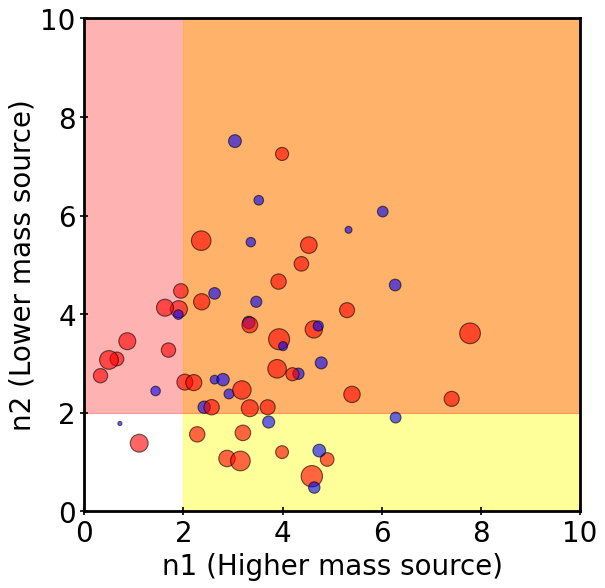}
        \caption{Combination of \sersic index in r-band}
        \label{fig:Sersic combination r-band}
    \end{subfigure}
    \hspace{0.4cm}
    \begin{subfigure}[b]{0.28\textwidth}
        \centering
        \includegraphics[width=\textwidth]{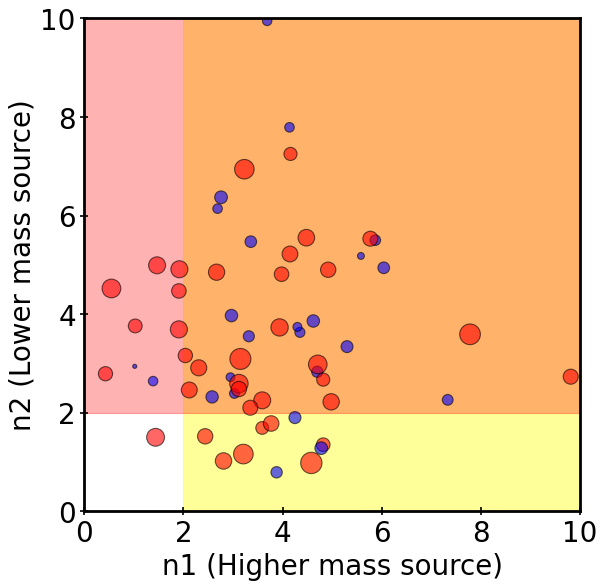}
        \caption{Combination of \sersic index in i-band}
        \label{fig:Sersic combination i-band}
    \end{subfigure}

    \caption{The 3 plots from left to right show, respectively, the pairing of \sersic index in merging samples, where $n_1$ and $n_2$ correspond to the \sersic index of the heavier and lighter bulge mass nuclei, respectively. The size of the scatter points depends inversely on the bulge mass ratio, where a lower bulge mass ratio ($1 < \frac{M_1}{M_2} < 3$) which is the major merger sample is indicated in red, and the higher bulge mass ratio ($\frac{M_1}{M_2} \geq 3$) which is the minor merger sample is represented as blue.}
    \label{sersic-combination}
\end{figure*}

The pairwise distribution of the \sersic indices for the 60 merging galaxies is shown in Figure~\ref{sersic-combination}. Triple systems are excluded for clarity, as the focus is on dual AGN systems. The plot is divided into four regions representing different bulge pair types: (i) classical–classical (C–C) pairs, which include mergers of elliptical galaxies or disc galaxies where both hosts have classical bulges. This region (orange) contains the largest number of bulges; (ii) classical–pseudo (C–P) pairs, spanning two regions (pink and yellow), include mergers between a classical and a pseudo-bulge host, either in disc or elliptical galaxies. In most cases, the classical bulge is more massive; (iii) pseudo–pseudo (P–P) pairs, forming the smallest group (white), and consist of mergers where both hosts have pseudo-bulges. Major mergers appear scattered across the diagram, while minor mergers are more concentrated in the C–C region. The bulge and host galaxy classifications for each source are listed in Table~\ref{table:properties}. 

One of the key questions addressed in this work is whether dual AGN are more frequently hosted by mergers of elliptical galaxies, disk galaxies, or mixed pairs. Table~\ref{table:properties} summarises the morphological distribution of dual AGN host galaxies, classified as elliptical–elliptical (E–E), disk–disk (D–D), and elliptical–disk (E–D) mergers. The respective counts are: E–E = 24, D–D = 18, and E–D = 24. Two systems are identified as triplets involving D–E–E and D–D–E configurations and are therefore counted twice. Although the sample is neither complete nor statistically unbiased, the results suggest that dual AGN are more commonly associated with elliptical (E–E) and mixed (E–D) mergers than with disk–disk (D–D) systems. This is consistent with the expectation that dual AGN are associated with galaxies that have undergone mergers, since elliptical galaxies preferentially reside in denser environments relative to disk galaxies \citep{dressler.1980}. A similar trend is evident in the colour–magnitude plot of the GOTHIC sample  \citep{Gothic2023}, where dual AGN predominantly occupy the red, evolved galaxy population in the plot.

Using GALFIT-derived scale radii, the bulge sizes of the 131 nuclei are compared. For disk galaxies, this corresponds to the bulge effective radius ($R_e$), while for ellipticals it represents the overall galaxy scale radius. Figure~\ref{fig:combined-plot}(a) shows that ellipticals exhibit $R_e$ values extending up to 21~kpc, whereas disk galaxies have $R_e < 6$kpc, which may correspond to a bar embedded within a faint disk or a disky elliptical galaxy. Overall, the plot highlights a clear dichotomy between the scale lengths of disks and those of elliptical galaxies.

In this study we consider mergers with galaxy mass ratios $\leq$3 to be major mergers and those with mass ratios $>$3 as minor mergers.  This threshold was chosen as a value close to the 1:1 ratio would be more accurate as a Major merger candidate and is also consistent with the numerical studies on mergers in the nearby universe \citep{conselice2006}.  Figure \ref{fig:combined-plot}(b) shows the distribution of major mergers and minor mergers.  One source appears to have a very large mass ratio ($1237655502962688659$). For this source one galaxy appears to be embedded inside the disk of the companion galaxy which appears to be relatively large. But overall our results indicate that $\sim$60\% of dual AGN are associated with major mergers. If we assume major mergers to have galaxy mass ratios $\leq$4, then the fraction of dual AGN in major mergers becomes even larger.

Finally, the bulge-to-total stellar mass ratio (B/T) was estimated for the sample of 58 disc galaxies. The distribution of B/T values, along with the bulge-to-disc ratio (B/D) as a function of total stellar mass, is presented in Figure~\ref{fig:combined-plot}(c). The B/T values range from $\sim$0.1 to 0.95. Systems with B/T~$>$~0.6 are bulge-dominated, exhibiting faint stellar discs. Such galaxies may correspond to S0 types or giant low surface brightness (GLSB) galaxies, both known to host diffuse stellar disks. As GLSB galaxies are typically isolated, the high B/T values are likely associated with S0 galaxies in the sample.

\begin{table*}
\begin{threeparttable}
\centering
\scriptsize
\caption{The table contains color values as obtained from Galfit output with further corrections such as extinction correction and K-correction. The color value obtained from this corrected magnitudes and the corresponding stellar mass calculated using the method mentioned in the text. \tnote{{\ddag}}}
\label{table: color correction}
\begin{tabular}{c*{16}{c}}
\hline
       & \multicolumn{4}{c}{BAND values} & \multicolumn{4}{c}{Ex corr.} & \multicolumn{4}{c}{K-Corr.} & \multicolumn{2}{c}{(g-r)}  & \multicolumn{2}{c}{Mass ($\times 10^{11} \textup{M}_\odot$}) \\ \cmidrule(lr){2-5}\cmidrule(lr){6-9} \cmidrule(lr){10-13} \cmidrule(lr){14-15} \cmidrule(lr){16-17} 
       objid  & \multicolumn{2}{c}{Bulge} & \multicolumn{2}{c}{Disk} & \multicolumn{2}{c}{Bulge} & \multicolumn{2}{c}{Disk} & \multicolumn{2}{c}{Bulge} & \multicolumn{2}{c}{Disk} & Bulge & disk & Bulge & disk\\ 
       \cmidrule(lr){2-3}\cmidrule(lr){4-5} \cmidrule(lr){6-7} \cmidrule(lr){8-9} \cmidrule(lr){10-11}\cmidrule(lr){12-13}
     & g & r & g & r & g & r & g & r & g & r & g & r &    &    &    &    \\
     \hline
             1237650762394959890 & 16.35 & 15.36 & ... & ... & 16.27 & 15.31 & ... & ... & 16.06 & 15.23 & ... & ... & 0.833 & ... & 4.277 & ... \\ 
        1237650762394959891 & 17.46 & 16.62 & ... & ... & 17.38 & 16.57 & ... & ... & 17.20 & 16.50 & ... & ... & 0.705 & ... & 0.973 & ... \\ \hline 
        1237651252018151487 & 16.75 & 15.8 & ... & ... & 16.62 & 15.71 & ... & ... & 16.43 & 15.64 & ... & ... & 0.794 & ... & 2.355 & ... \\ 
        1237651252018151484 & 16.28 & 15.26 & ... & ... & 16.15 & 15.17 & ... & ... & 15.96 & 15.10 & ... & ... & 0.860 & ... & 4.224 & ... \\ \hline 
        1237652600110383328 & 18.17 & 17.36 & ... & ... & 18.01 & 17.25 & ... & ... & 17.83 & 17.23 & ... & ... & 0.596 & ... & 1.499 & ... \\ 
        1237652600110383327 & 18.81 & 17.9 & ... & ... & 18.65 & 17.79 & ... & ... & 18.38 & 17.73 & ... & ... & 0.653 & ... & 1.383 & ... \\ \hline 
        1237653441374453925 & 20.06 & 18.54 & 16.41 & 15.6 & 19.96 & 18.47 & 16.31 & 15.53 & 20.10 & 18.49 & 16.16 & 15.47 & 1.609 & 0.688 & 0.960 & 1.860 \\ 
        1237653441374453924 & 17.27 & 16.37 & ... & ... & 17.17 & 16.30 & ... & ... & 17.00 & 16.24 & ... & ... & 0.768 & ... & 1.092 & ... \\ \hline 
        1237654382516240489 & 18.43 & 17.13 & 17.49 & 16.47 & 18.30 & 17.04 & 17.36 & 16.38 & 17.91 & 16.90 & 17.04 & 16.27 & 1.008 & 0.769 & 3.601 & 3.102 \\ 
        1237654382516240490 & 18.81 & 17.75 & ... & ... & 18.68 & 17.66 & ... & ... & 18.35 & 17.55 & ... & ... & 0.801 & ... & 1.247 & ... \\ \hline 

        \end{tabular}

        \begin{tablenotes}
\item[{\ddag}] This is sample table with limited elements and the entire table is available in electronic format.
\end{tablenotes}
\end{threeparttable}
        \end{table*}

\begin{table*}
\centering
\begin{threeparttable}
\caption{A table representing the merging galaxy pairs. The combinations are Galaxy type (Ellipticals or Disky); Bulge type (Classical or Pseudo); Mass Ratio (Major merger if mass ratios $<$ 3, Minor merger for values $>$ 3) \tnote{{\dag}}}
\label{table:properties}
\begin{tabular}{ccccccccc}
\hline
        No. & objid & Exp &  disc & Sersic & bulge &  bulge & stellar & Mass\\
            &       &  disc &  combination  & index & type & pair & mass & ratio\\
        \hline
1 & 1237650762394959890 & No & \multirow{2}{*}{Elip-Elip} & 5.87 & Classical & \multirow{2}{*}{C-C} & 4.277 & \multirow{2}{*}{4.396} \\
1 & 1237650762394959891 & No & ~ & 5.5 & Classical & ~ & 0.973 & ~ \\\hline

2 & 1237651252018151487 & No & \multirow{2}{*}{Elip-Elip} & 5.55 & Classical & \multirow{2}{*}{C-C} & 2.355 & \multirow{2}{*}{1.793} \\
2 & 1237651252018151484 & No & ~ & 4.48 & Classical & ~ & 4.224 & ~ \\\hline

3 & 1237652600110383328 & No & \multirow{2}{*}{Elip-Elip} & 4.58 & Classical & \multirow{2}{*}{C-P} & 1.499 & \multirow{2}{*}{1.084} \\
3 & 1237652600110383327 & No & ~ & 0.98 & Pseudo & ~ & 1.383 & ~ \\\hline

4 & 1237653441374453925 & Yes & \multirow{2}{*}{ disc-Elip} & 1.03 & Pseudo & \multirow{2}{*}{C-P} & 2.820 & \multirow{2}{*}{2.582} \\
4 & 1237653441374453924 & No & ~ & 3.76 & Classical & ~ & 1.092 & ~ \\\hline

5 & 1237654382516240489 & Yes & \multirow{2}{*}{ disc-Elip} & 4.14 & Classical & \multirow{2}{*}{C-C} & 6.702 & \multirow{2}{*}{5.377} \\
5 & 1237654382516240490 & No & ~ & 7.79 & Classical & ~ & 1.247 & ~ \\\hline
\end{tabular}
\begin{tablenotes}
\item[{\dag}] This is sample table with limited elements and the entire table is available in electronic format.
\end{tablenotes}
\end{threeparttable}
\end{table*}

\begin{table}
\centering
\caption{Bulge Type vs. Disk Presence}
\label{tab:bulge_disk}
\begin{tabular}{|l|c|c|c|}
\hline
\textbf{} & \textbf{Classical Bulge} & \textbf{Pseudo Bulge} & \textbf{Total} \\
\hline
\textbf{No Disk} & 67 & 6 & 73 \\
\textbf{With Disk} & 38 & 20 & 58 \\
\hline
\textbf{Total} & 105 & 26 & 131 \\
\hline
\end{tabular}
\end{table}

\begin{figure*}
\centering
    \includegraphics[width=\textwidth]{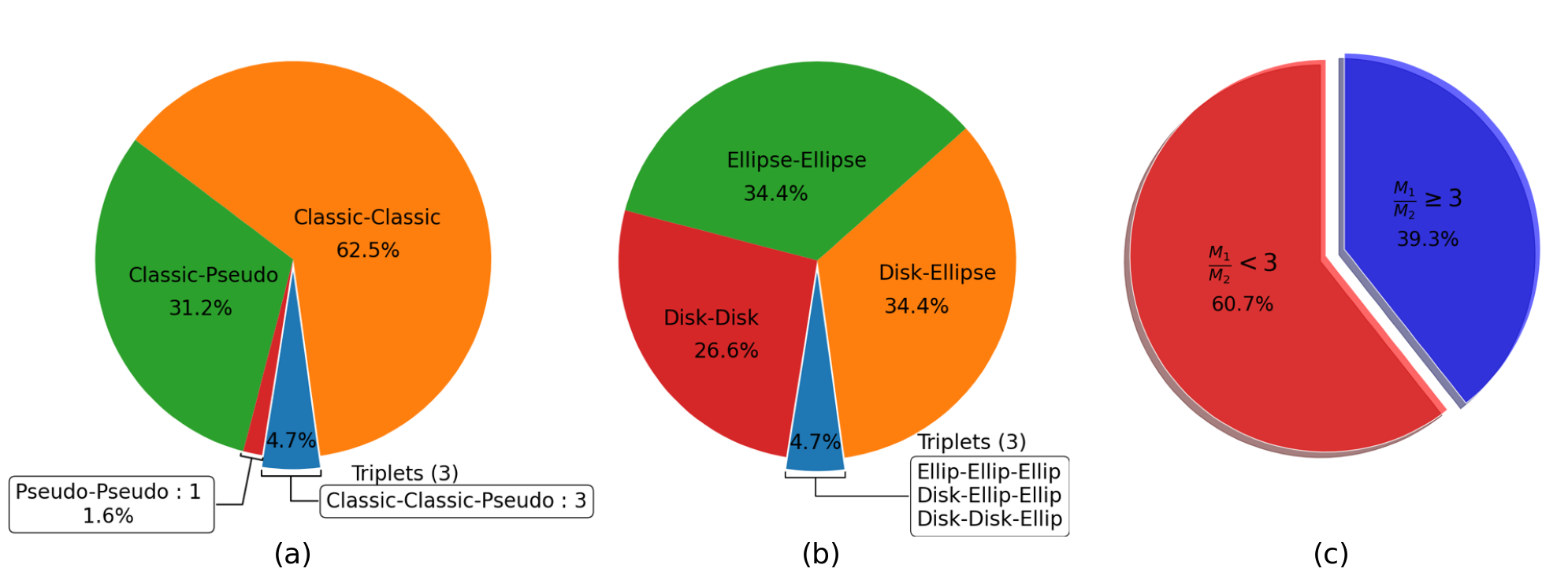}
        \caption{Pie diagrams from left to right. (a) Extreme left shows the different types of bulge combinations. (b) Center shows the fraction of different merger types. (c) The extreme right shows the major vs minor merger numbers.}
    \label{fig:sersic combination}
\end{figure*}

\begin{figure*}
    \centering
    \includegraphics[width=5cm]{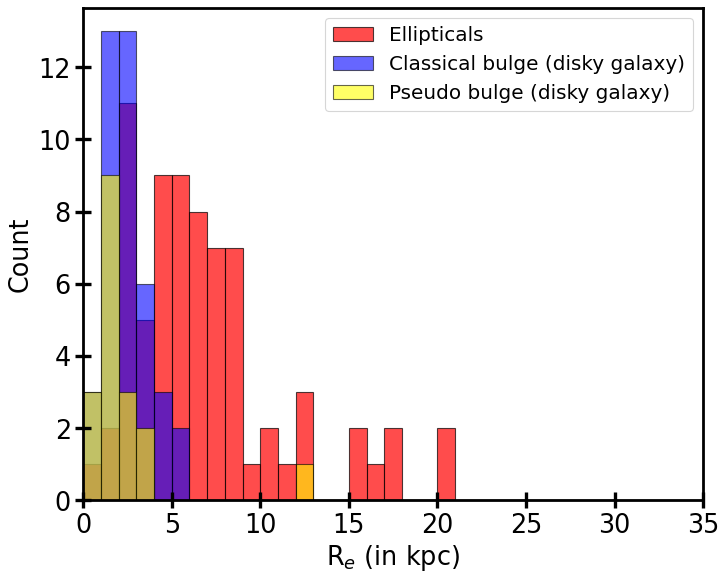}
    \hspace{0.5cm}
    \includegraphics[width=5.2cm]{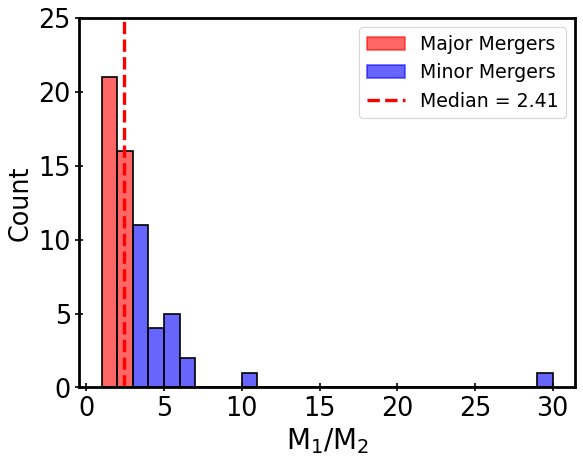}
    \hspace{0.5cm}
    \includegraphics[width=4.9cm]{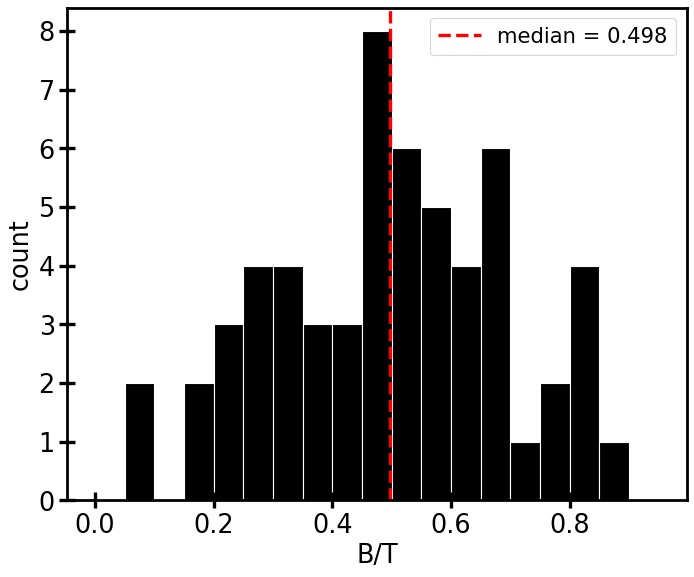}
    \caption{From left to right. (a) The plot shows the distribution of bulge scale radius $R_e$ in disk galaxies or the galaxy scale radii of elliptical galaxies. (b) The plot shows the distribution of total galaxy mass ratios for the sample. Note that there are 2 outliers having mass ratios $\sim$10 and $\sim$30. (c) The ratio of bulge to total galaxy mass for the 58 disk galaxies in our sample, where the median value is shown with a dashed red vertical line.}
    \label{fig:combined-plot}
\end{figure*}

\section{Discussion}
The structural and morphological analysis of dual AGN host galaxies offers important insights into the conditions conducive to forming and detecting supermassive black hole SMBH pairs. Our results reveal a strong preference for dual AGN to reside in systems with classical bulges, typically characterised by high \sersic\ indices ($n > 2$) and elevated bulge masses. This preference is closely tied to the evolutionary history of their host galaxies, with major mergers playing a central role in shaping the stellar and dynamical environments favourable for dual AGN activity. In the following paragraphs we examine the key trends emerging from our study and place it in the broader context of AGN triggering mechanisms and SMBH binary formation.

Our structural analysis reveals that dual AGN are predominantly hosted by galaxies with \sersic\ indices $n > 2$, indicating the presence of classical bulges that are typically found in elliptical or bulge-dominated disk galaxies. As shown in Figure~\ref{fig:sersic_index_distribution}, the median \sersic\ index across all wavebands exceeds 2. So classical bulges are notably prevalent among dual AGN host galaxies. Classical bulges also exhibit higher bulge masses (Figures~\ref{fig:bulge mass dist} and \ref{fig:mosaic-bulge-disk}). Assuming that the established $M$–$\sigma$ relation observed in AGN host galaxies is valid for the bulges of dual AGN \citep{mcconnel.ma.2013}, our results suggest that the merging process leads to the growth of SMBHs as well. It also suggests that SMBH binaries preferentially reside in merger remnants with $n > 2$.

Dual AGN are also more commonly found in elliptical pairs or elliptical–disk pairs. Based on \sersic\ model fits (Section 3.5), systems with a single high-\sersic\ component are classified as ellipticals, while those with two components are identified as bulge–disk systems. Assuming this classification, approximately two-thirds of the sample contain classical bulges in both galaxies (Figure~\ref{fig:sersic combination}). Also, 94\% of of dual AGN host at least one such bulge (Figures~\ref{sersic-combination} and \ref{fig:sersic combination}). As mentioned earlier, classical bulges are generally formed from the monolithic collapse of galaxies at early epochs or via the major mergers of galaxies. They are characterized by redder, older stellar populations, in contrast to pseudobulges that are formed via secular processes and are relatively bluer in color \citep{hu.etal.2024}. This trend is consistent with previous findings \citep{Gothic2023}, which show that dual AGN preferentially occupy the red sequence in colour–magnitude space.

Pseudobulges are rare among dual AGN hosts, and systems comprising of two pseudobulges are exceptionally uncommon (Figure~\ref{fig:sersic combination}). This further reinforces the link between classical bulge formation via mergers and the occurrence of dual AGN, as opposed to dual AGN in pseudobulges formed through secular evolution.

Finally, dual AGN are predominantly associated with major mergers, with typical stellar mass ratios $\leq 3$, as seen in the final panel of Figure~\ref{fig:sersic combination}. This result is consistent with earlier observational studies \citep{stemo.etal.2021} and suggests that major mergers play a key role not only in AGN triggering but also in the formation of SMBH binaries. These findings have important implications for the identification and characterisation of SMBH binaries in forthcoming low-redshift surveys.

\section{Conclusions}
In this study, we investigate the structural properties, host galaxy morphologies, and bulge classifications of dual AGN systems in the GOTHIC sample, based on two-dimensional decompositions of SDSS imaging. Our main results are as follows:
\begin{enumerate}
\item The bulge masses of dual AGN host galaxies span $1.5\times10^{9}$ to $1.4\times10^{12},M_\odot$, with the most massive bulges corresponding to elliptical galaxies. 

\item \sersic indices in the g, r, and i bands indicate that 80\% (105/131) of bulges are classical. Of these, 64\% (67) are found in elliptical galaxies, while 36\% (38) reside in disk galaxies. The remaining 20\% (26) are pseudobulges with significant disky components. 

\item Host morphologies reveal that dual AGN are more frequently found in elliptical–elliptical (E–E) and elliptical–disk (E–D) mergers than in disk–disk (D–D) mergers. Approximately two-thirds of the systems include at least one elliptical galaxy, while only ~30\% involve two disk galaxies. This suggests a preference for red, evolved hosts over star-forming systems. 

\item Nearly 60\% of dual AGN are associated with major mergers. Combined with their prevalence in red, quiescent galaxies, this supports the interpretation that dual AGN predominantly reside in quenched environments with low star formation activity. 
\end{enumerate}
\noindent
These findings suggest that dual AGN predominantly arise from major mergers and are preferentially hosted by massive, evolved systems—either classical bulges in disks or elliptical galaxies—consistent with merger-driven evolutionary pathways.


\section*{Acknowledgements}
MD and SB acknowledge the support of the Science and Engineering Research Board (SERB) Core Research Grant CRG/2022/004531 and the Department of Science and Technology (DST) grant DST/WIDUSHIA/PM/2023/25(G) for this research. This research has used SDSS data. Funding for the SDSS and SDSS-II have been provided by the Alfred P. Sloan Foundation, the Participating Institutions, the National Sr cience Foundation, the U.S. Department of Energy, the National Aeronautics and Space Administration, the Japanese Monbukagakusho, the Max Planck Society, and the Higher Education Funding Council for England. The SDSS Web Site is http://www.sdss.org/. The SDSS is managed by the Astrophysical Research Consortium for the Participating Institutions. The Participating Institutions are the American Museum of Natural History, Astrophysical Institute Potsdam, University of Basel, University of Cambridge, Case Western Reserve University, University of Chicago, Drexel University, Fermilab, the Institute for Advanced Study, the Japan Participation Group, Johns Hopkins University, the Joint Institute for Nuclear Astrophysics, the Kavli Institute for Particle Astrophysics and Cosmology, the Korean Scientist Group, the Chinese Academy of Sciences (LAMOST), Los Alamos National Laboratory, the MaxPlanck-Institut fur¨ Astronomie (MPIA), the Max-Planck-Institut fuAstrophysik (MPA), New Mexico State University, Ohio State University, University of Pittsburgh, University of Portsmouth, Princeton University, the United States Naval Observatory, and the University of Washington.

\section*{Data Availability}
This article has made use SDSS DR18 data which is public data and available online. 
 


\bibliographystyle{mnras}
\bibliography{ref} 









\section{Supporting Information }
Supplementary data from Table 1 and Table 2 are available at MNRAS online.
 



\bsp	
\label{lastpage}
\end{document}